\documentclass[runningheads]{llncs}
\usepackage[T1]{fontenc}
\usepackage{graphicx}
\usepackage{amsfonts}
\usepackage{tikz}
\usetikzlibrary{fit,backgrounds,positioning}
\usepackage{subcaption}
\usepackage{bm}

\DeclareCaptionFormat{cont}{#1 continued#2#3\par}

\newcommand{\add}{\ensuremath{\textsc{Add}}}
\newcommand{\remove}{\ensuremath{\textsc{Remove}}}
\newcommand{\sumq}{\ensuremath{\textsc{Sum}}}
\newcommand{\topk}{\ensuremath{\textsc{Top}}}

\newcommand{\dist}{\ensuremath{\mathrm{dist}}}
\newcommand{\RR}{\mathbb{R}}

\usepackage{cleveref}
\crefname{theorem}{Theorem}{Theorems}
\crefname{section}{Section}{Sections}
\crefname{appendix}{Appendix}{Appendices}
\crefname{figure}{Fig.}{Figs.}

\begin{document}


\title{Sum-of-Local-Effects Data Structures\\for Separable
Graphs\thanks{Supported by NSERC Discovery Grants RGPIN-07185-2020 and RGPIN-05435-2018.}}
\titlerunning{SOLE Data Structures}

\author{Xing Lyu\inst{1} \and
Travis Gagie\inst{2}\orcidID{0000-0003-3689-327X} \and
Meng He\inst{2}\orcidID{0000-0003-0358-7102} \and
Yakov Nekrich\inst{3}\orcidID{0000-0003-3771-5088} \and
Norbert Zeh\inst{2}\orcidID{0000-0002--0562-1629}}
\authorrunning{X. Lyu et al.}

\institute{Halifax West High School, Halifax, Canada \email{lyuxing1006@gmail.com} \and
Dalhousie University, Halifax, Canada \email{\{firstname.lastname\}@dal.ca} \and
Michigan Technological University \email{yakov@mtu.edu}}

\maketitle


\begin{abstract}
  It is not difficult to think of applications that can be modelled as graph
  problems in which placing some facility or commodity at a vertex has some
  positive or negative effect on the values of all the vertices out to some
  distance, and we want to be able to calculate quickly the cumulative effect on
  any vertex's value at any time or the list of the most beneficial or most
  detrimential effects on a vertex.  In this paper we show how, given an
  edge-weighted graph with constant-size separators, we can
  support the following operations on it in time polylogarithmic in the number
  of vertices and the number of facilities placed on the vertices, where
  distances between vertices are measured with respect to the edge weights:
  \begin{description}
    \item[\boldmath $\add (v, f, w, d)$] places a facility of weight $w$ and
    with effect radius $d$ onto vertex $v$.
    \item[\boldmath $\remove (v, f)$] removes a facility $f$ previously placed
    on $v$ using $\add$ from $v$.
    \item[\boldmath $\sumq (v)$ or $\sumq(v, d)$] returns the total weight of
    all facilities affecting $v$ or, with a distance parameter $d$, the total
    weight of all facilities whose effect region intersects the ``circle'' with
    radius $d$ around $v$.
    \item[\boldmath $\topk(v, k)$ or $\topk(v, k, d)$] returns the $k$ facilities
    of greatest weight that affect $v$ or, with a distance parameter $d$, whose
    effect region intersects the ``circle'' with radius $d$ around $v$.
  \end{description}
  The weights of the facilities and the operation that $\sumq$ uses to ``sum''
  them must form a semigroup. For $\topk$ queries, the weights must be drawn from
  a total order.

\end{abstract}

\section{Introduction}

\label{sec:introduction}

Even people who have never heard of Baron Samuel of Wych Cross may have heard a
saying often attributed to him, that there are three things that matter in real
estate: location, location, location.  This means that the value of a property
may increase or decrease depending on whether it is close to a bus stop, a good
school, a supermarket or a landfill, for example.  Of course, ``close'' may not
mean the same thing for a bus stop as it does for a landfill, and the positive
effect of the former may not offset the negative effect of the latter.  In fact,
``close'' may not refer to Euclidean distance, since walking to a bus stop five
minutes down the street is preferable to walking to one five minutes away
through a landfill.  To model applications in which there are such additive
local effects with a non-Euclidean definition of locality, we propose in this
paper a data structure for a graph $G$ that supports the following operations:
\begin{description}
  \item[\boldmath $\add (v, f, w, d)$] places a facility of weight $w$ onto
  vertex $v$. The \emph{effect region} of $f$ is a circle with radius $d$ around
  $v$.
  \item[\boldmath $\remove (v, f)$] removes a facility $f$ previously placed
  on $v$ using $\add$ from $v$.
  \item[\boldmath $\sumq (v)$ or $\sumq(v, d)$] returns the total weight of
  all facilities affecting $v$ or, with a distance parameter $d$, the total
  weight of all facilities whose effect region intersects the ``circle'' with
  radius $d$ around $v$.
  \item[\boldmath $\topk(v, k)$ or $\topk(v, k, d)$] returns the $k$ facilities
  of greatest weight that affect $v$ or, with a distance parameter $d$, whose
  effect region intersects the ``circle'' with radius $d$ around $v$.
\end{description}
We assume that every edge $e \in G$ has a non-negative length $\ell(e)$ and that
distances between vertices are measured as the minimum total length of all
edges on any path between these two vertices.  A circle with radius $d$ around
some vertex $v$ includes all vertices and (parts of) edges at distance $d$ from
$v$.  More precisely, if $f$ is a facility with effect radius $d'$ placed on
some vertex $u$, then we consider $f$'s effect region to intersect a circle with
radius $d$ around some other vertex $v$ if and only if $\dist(u,v) \le d + d'$.

The weights of the facilities and the operation that $\sumq$ uses to ``sum''
them must form a semigroup (see \cref{sec:semigroup}). For $\topk$ queries, the
weights must be drawn from a total order. Note that $\sumq(v)$ and $\topk(v, k)$
can be viewed as ``range stabbing queries on graphs'', whereas $\sumq(v, d)$ and
$\topk(v, k, d)$ with $d > 0$ are ``range intersection queries on graphs,''
where the ranges are the effect regions of the facilities and a query is either
an individual vertex or a region of some radius $d$ around some vertex.
  
We call such a data structure a sum-of-local-effects (SOLE) data structure.  In
Section~\ref{sec:trees}, we show that when $G$ is a tree on $n$ vertices, then
there is a SOLE data structure for it supporting $\add$, $\remove$, and $\sumq$
operations in $O (\lg n\lg m)$ time, and $\topk$ queries in $O(k \lg n \lg m)$
time, where $m$ is the total number of facilities currently placed on the
vertices of $G$.  In Section~\ref{sec:sp_graphs}, we generalize this result to
$t$-separable graphs, for any constant $t$, which includes series-parallel
graphs $(t \le 2$), graphs of constant treewidth $w$ ($t \le w + 1$), and graphs
of constant branchwidth $b$ ($t \le b$).  We show that when $G$ is
$t$-separable, there exists a SOLE data structure for it supporting $\add$,
$\remove$, and $\sumq$ operations in $O (\lg n \lg^t m)$ time, and $\topk$
queries in $O(k \lg n\lg^t m)$ time.  The consts of $\add$ and $\remove$
operations are amortized in this case.

We believe that all our results can be extended to \emph{directed} graphs $G$
and that our data structure can be made to support vertex and edge deletions in
$G$.  We will investigate this generalization in a future version of this paper.

\section{A SOLE Data Structure for Trees}

\label{sec:trees}

In this section, we prove that

\begin{theorem}
  \label{thm:trees}
  If $G$ is a tree on $n$ vertices, then there is a SOLE data structure for it
  supporting $\add$, $\remove$, and $\sumq$ operations in $O (\lg n \lg m)$
  time, and $\topk(v, k, d)$ operations in $O(k \lg n \lg m)$ time, where $m$ is
  the number of facilities currently on the vertices of $G$. The size of
  this data structure is $O(n + m \lg n)$.
\end{theorem}

Assume for now that all vertices of $G$ have degree at most $3$.  We choose an
arbitrary vertex $\rho$ of $G$ and label every vertex $v$ in $G$ with its
distance $\dist(\rho, v)$ from $\rho$. A \emph{centroid edge} of $G$ is an edge
$(u,v)$ whose removal splits $G$ into two subtrees $G_u$ and $G_v$ with at most
$2n/3$ vertices each. Such an edge exists because all vertices of $G$ have
degree at most $3$.  A \emph{centroid decomposition} of $G$ is a binary tree $T$
defined inductively as follows (see \cref{fig:centroid}): If $G$ has a single
vertex $v$, then $T$ has $v$ as its only node. Otherwise, let $(u,v)$ be an
arbitrary centroid edge of $G$.  Then the root of $T$ is $(u,v)$, and the two
children of $(u,v)$ are the roots of centroid decompositions of $G_u$ and $G_v$.
For each edge $e$ of $T$, let $T_e$ be the subtree of $T$ below $e$, and let
$V_e$ be the set of vertices of $G$ corresponding to the leaves of $T_e$. The
height of $T$ is at most $\log_{3/2}n = O(\lg n)$.

Our SOLE data structure for $G$ consists of a centroid decomposition $T$ of~$G$
where each edge $e$ of $T$ has an associated data structure $W_e$ storing
facilities that may affect the vertices in $V_e$. Each leaf $v$ of $T$
(corresponding to the vertex $v$ of $G$) also has an associated data structure
$W_v$ storing the facilities placed on $v$ itself.  Each facility $f$ with
weight $w$ in $W_e$ or $W_v$ has an associated \emph{radius} $r$ and is stored
as the triple $(r, f, w)$ in $W_e$.

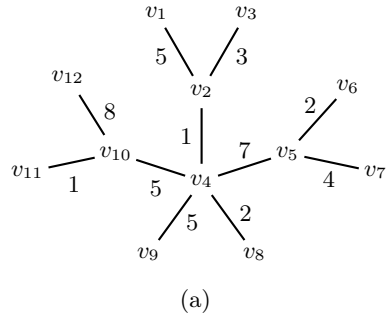
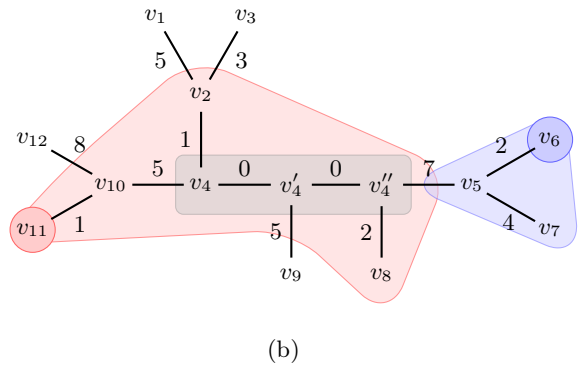
\begin{figure}[p]
  \centering
  \subfloat[\label{fig:tree}]{%
    \begin{tikzpicture}[
      vertex/.style={circle,inner sep=1pt},
      edge/.style={thick}
    ]
      \path           node [vertex] (v1)  {$v_{1}$}
      ++(300:1.2)     node [vertex] (v2)  {$v_{2}$}
      +(60:1.2)       node [vertex] (v3)  {$v_{3}$}
      ++(270:1.2)     node [vertex] (v4)  {$v_{4}$}
      ++(18:1.2)      node [vertex] (v5)  {$v_{5}$}
      +(48:1.2)       node [vertex] (v6)  {$v_{6}$}
      +(348:1.2)      node [vertex] (v7)  {$v_{7}$}
      (v4) +(306:1.2) node [vertex] (v8)  {$v_{8}$}
      +(234:1.2)      node [vertex] (v9)  {$v_{9}$}
      ++(162:1.2)     node [vertex] (v10) {$v_{10}$}
      +(192:1.2)      node [vertex] (v11) {$v_{11}$}
      +(122:1.2)      node [vertex] (v12) {$v_{12}$};
      \draw [edge]
      (v1)  to node [pos=0.6,xshift=-2pt,left]   {$5$} (v2)
      (v2)  to node [pos=0.4,xshift=2pt,right]   {$3$} (v3)
      (v2)  to node [left]                       {$1$} (v4)
      (v4)  to node [above]                      {$7$} (v5)
      (v5)  to node [pos=0.3,above,yshift=2pt]   {$2$} (v6)
      (v5)  to node [pos=0.45,below]             {$4$} (v7)
      (v4)  to node [pos=0.4,right,xshift=2pt]   {$2$} (v8)
      (v4)  to node [pos=0.6,right,xshift=2pt]   {$5$} (v9)
      (v4)  to node [pos=0.6,below,yshift=-2pt]  {$5$} (v10)
      (v10) to node [pos=0.45,below,yshift=-2pt] {$1$} (v11)
      (v10) to node [pos=0.6,right,xshift=2pt]   {$8$} (v12);
    \end{tikzpicture}
  }

  \bigskip
  \bigskip

  \subfloat[\label{fig:degree-3}]{%
    \begin{tikzpicture}[
      vertex/.style={circle,inner sep=1pt},
      edge/.style={thick}
    ]
      \path            node [vertex] (v1)  {$v_{1}$}
      ++(300:1.2)      node [vertex] (v2)  {$v_{2}$}
      +(60:1.2)        node [vertex] (v3)  {$v_{3}$}
      ++(270:1.2)      node [vertex] (v4) {$v_{4}$}
      ++(0:1.2)        node [vertex] (v4a) {$v_{4}'$}
      ++(0:1.2)        node [vertex] (v4b) {$v_{4}''$}
      ++(0:1.2)        node [vertex] (v5)  {$v_{5}$}
      +(30:1.2)        node [vertex] (v6)  {$v_{6}$}
      +(330:1.2)       node [vertex] (v7)  {$v_{7}$}
      (v4b) +(270:1.2) node [vertex] (v8)  {$v_{8}$}
      (v4a) +(270:1.2) node [vertex] (v9)  {$v_{9}$}
      (v4) ++(180:1.2) node [vertex] (v10) {$v_{10}$}
      +(210:1.2)       node [vertex] (v11) {$v_{11}$}
      +(150:1.2)       node [vertex] (v12) {$v_{12}$};
      \draw [edge]
      (v1)  to node [pos=0.6,xshift=-2pt,left]  {$5$} (v2)
      (v2)  to node [pos=0.4,xshift=2pt,right]  {$3$} (v3)
      (v2)  to node [left]                      {$1$} (v4)
      (v4b) to node [above]                     {$7$} (v5)
      (v5)  to node [pos=0.3,above,yshift=2pt]  {$2$} (v6)
      (v5)  to node [pos=0.45,below]            {$4$} (v7)
      (v4b) to node [left]                      {$2$} (v8)
      (v4a) to node [left]                      {$5$} (v9)
      (v4)  to node [above]                     {$5$} (v10)
      (v10) to node [pos=0.3,below,yshift=-2pt] {$1$} (v11)
      (v10) to node [pos=0.3,above,yshift=2pt]  {$8$} (v12)
      (v4)  to node [above]                     {$0$} (v4a)
      (v4a) to node [above]                     {$0$} (v4b);
      \begin{scope}[on background layer]
        \coordinate (1) at ([xshift=-5mm,yshift=-2mm]v11.center);
        \coordinate (2) at (barycentric cs:v10=0.5,v12=0.5);
        \coordinate (3) at ([xshift=-1mm,yshift=5mm]v2.center);
        \coordinate (5) at ([yshift=2mm]barycentric cs:v4b=0.25,v5=0.75);
        \coordinate (6) at ([yshift=-6mm,xshift=1mm]v8.center);
        \coordinate (7) at (barycentric cs:v4a=0.5,v9=0.5);
        \path [rounded corners=5mm,draw=red!80,fill=red!20,opacity=0.5] (1) -- (2) -- (3) -- (5) -- (6) -- (7) -- cycle;
        \coordinate (1) at ([xshift=2mm,yshift=4mm]v6.center);
        \coordinate (2) at ([xshift=4mm,yshift=-4mm]v7.center);
        \coordinate (3) at ([xshift=-3mm]barycentric cs:v4b=0.5,v5=0.5);
        \path [rounded corners=5mm,draw=blue!60,fill=blue!20,opacity=0.5] (1) -- (2) -- (3) -- cycle;
        \node [draw=red!60,fill=red!20,circle,inner sep=0pt,minimum size=6mm] at (v11) {};
        \node [draw=blue!60,fill=blue!20,circle,inner sep=0pt,minimum size=6mm] at (v6) {};
        \node [draw=black!40,fill=black!20,opacity=0.5,fit={(v4) (v4a) (v4b)},rounded corners=3pt] {};
      \end{scope}
    \end{tikzpicture}
  }

  \caption{A tree (\subref{fig:tree}), its degree-3 version
  (\subref{fig:degree-3}), and the centroid decomposition
  (\subref{fig:centroid}, next page) of the tree in (\subref{fig:degree-3}). The
  shaded subtree in (\subref{fig:degree-3}) is the degree-3 tree replacing the
  high-degree vertex $v_4$ in (\subref{fig:tree}). A facility $f$ with effect
  radius 8 placed on $v_{11}$ has the pink effect region in
  (\subref{fig:degree-3}).  This region overlaps the blue query region with
  radius 8 around $v_6$. In the centroid decomposition (\subref{fig:centroid}),
  $f$ is stored in the edge data structures of the fat red edges and the node
  data structure of $v_{11}$, with the radii shown in red. A query with radius
  $8$ around $v_6$ queries the node data structure of $v_6$ an the edge data
  structures of the fat blue edges, with the query radii shown in blue. In
  particular, $f$ is reported as part of the query on the data structure $W_e$
  associated with the highlighted child edge of $(v_4, v_4')$.}
\end{figure}
\begin{figure}[t]
  \centering\ContinuedFloat
  \subfloat[\label{fig:centroid}]{%
    \begin{tikzpicture}[
      vertex/.style={circle,inner sep=1pt},
      edge/.style={thick}
    ]
    \path node (v2) {$v_2$}
    node [right=4mm of v2]     (v3)     {$v_3$}
    node [right=4mm of v3]     (v10v12) {$(v_{10},v_{12})$}
    node [right=4mm of v10v12] (v11)    {$v_{11}$}
    node [right=4mm of v11]    (v4a)    {$v_4'$}
    node [right=4mm of v4a]    (v9)     {$v_9$}
    node [right=4mm of v9]     (v4b)    {$v_4''$}
    node [right=4mm of v4b]    (v8)     {$v_8$}
    node [right=4mm of v8]     (v5)     {$v_5$}
    node [right=4mm of v5]     (v7)     {$v_7$}
    node [anchor=north east,shift={(-2mm,-8mm)}] at (v10v12.south) (v10) {$v_{10}$}
    node [anchor=north west,shift={(2mm,-8mm)}]  at (v10v12.south) (v12) {$v_{12}$}
    ({barycentric cs:v10v12=0.5,v11=0.5} |- v10v12.north) node [anchor=south,yshift=8mm] (v10v11) {$(v_{10},v_{11})$}
    node [left=4mm of v10v11]  (v4)     {$v_4$}
    ({barycentric cs:v2=0.5,v3=0.5} |- v10v11) node (v2v3) {$(v_2,v_3)$}
    node [left=4mm of v2v3]    (v1)     {$v_1$}
    ({barycentric cs:v4a=0.5,v9=0.5} |- v10v11) node (v4av9) {$(v_4',v_9)$}
    ({barycentric cs:v4b=0.5,v8=0.5} |- v10v11) node (v4bv8) {$(v_4'',v_8)$}
    ({barycentric cs:v5=0.5,v7=0.5} |- v10v11)  node (v5v7)  {$(v_5,v_7)$}
    node [right=4mm of v5v7]   (v6)     {$v_6$}
    ({barycentric cs:v10v11=0.5,v4=0.5} |- v10v11.north) node [anchor=south,yshift=8mm] (v4v10) {$(v_4,v_{10})$}
    ({barycentric cs:v1=0.5,v2v3=0.5} |- v4v10)  node (v1v2)  {$(v_1,v_2)$}
    ({barycentric cs:v4av9=0.5,v4bv8=0.5} |- v4v10)  node (v4av4b)  {$(v_4',v_4'')$}
    ({barycentric cs:v5v7=0.5,v6=0.5} |- v4v10)  node (v5v6)  {$(v_5,v_6)$}
    ({barycentric cs:v1v2=0.5,v4v10=0.5} |- v4v10.north) node [anchor=south,yshift=8mm] (v2v4) {$(v_2,v_4)$}
    ({barycentric cs:v4av4b=0.5,v5v6=0.5} |- v2v4)  node (v4bv5)  {$(v_4'',v_5)$}
    ({barycentric cs:v2v4=0.5,v4bv5=0.5} |- v4bv5.north) node [anchor=south,yshift=8mm] (v4v4a) {$(v_4,v_4')$};
    \draw [edge] (v2) -- (v2v3) -- (v3)
    (v1) -- (v1v2) -- (v2v3)
    (v10) -- (v10v12) -- (v12)
    (v4v4a) -- (v2v4) -- (v4v10) -- (v10v11) -- (v11)
    (v4a) -- (v4av9) -- (v9)
    (v4b) -- (v4bv8) -- (v8)
    (v4av9) -- (v4av4b) -- (v4bv8)
    (v5) -- (v5v7) -- (v7)
    (v5v7) -- (v5v6)
    (v4av4b) -- (v4bv5)
    ;
    \draw [ultra thick,red] (v2v4) to node [left,yshift=2pt,pos=0.4] {$1$} (v1v2)
    (v4v10) to node [left,pos=0.4] {$2$} (v4)
    (v10v11) to node [left,pos=0.4] {$7$} (v10v12);
    \begin{scope}
      \path (v4v4a.center) +(225:15pt) coordinate (start);
      \clip (start) -- (v4v4a.center) -- (v4bv5.center) -- +(225:15pt) -- cycle;
      \draw [line width=3.2pt,red] (v4v4a) to node [below,pos=0.45] {$2$} (v4bv5);
    \end{scope}
    \draw [ultra thick,blue] (v4bv5) to node [right,xshift=3pt,pos=0.35] {$-6$} (v5v6)
    (v5v6) to node [right,pos=0.35] {$-8$} (v6);
    \begin{scope}
      \path (v4v4a.center) +(0:30pt) coordinate (start);
      \clip (start) -- (v4v4a.center) -- (v4bv5.center) -- +(90:15pt) -- cycle;
      \draw [line width=3.2pt,blue] (v4v4a) to node [above,pos=0.5] {$1$} (v4bv5);
    \end{scope}
    \begin{scope}[on background layer]
      \node [draw=red!60,fill=red!20,circle,inner sep=0pt,minimum size=6mm] at (v11) {};
      \node [red,anchor=north,yshift=-0.75mm] at (v11.south) {$8$};
      \node [draw=blue!60,fill=blue!20,circle,inner sep=0pt,minimum size=6mm] at (v6) {};
    \end{scope}
    \end{tikzpicture}
  }

  \captionsetup{format=cont}
  \caption{}
\end{figure}

We represent each data structure $W_x$, where $x$ can be an edge or a leaf of
$T$, as two search trees $R_x$ and $F_x$.  $R_x$ is a priority search tree
\cite{mccreightPrioritySearchTrees1985} (see \cref{sec:priority-search-tree}) on
the triples $(r, f, w)$ in~$W_x$, using the radii $r$ as $x$-coordinates and the
weights $w$ as $y$-coordinates. Each node $v$ of $R_x$ is augmented with the
total weight of all triples in the subtree below $v$ (see
\cref{sec:range-sum-ds}). We call this combination of a priority search tree and
a range sum data structure a ``range sum priority search tree'' in
\cref{sec:combinations}.  $F_x$ is a standard search tree over the triples $(r,
f, w)$ in $W_x$, using the identifiers $f$ of facilities as keys.  The two
copies of $(r,f,w)$ in $R_x$ and $F_x$ are linked using cross pointers.  Thus,
$W_x$ supports the following operations in $O(\lg m)$ time: insertion of a new
triple $(r,f,w)$, deletion of a triple associated with facility $f$, and
reporting of the total weight of all triples $(r,f,w)$ with $q \le r$, for some
query radius $q$. It also supports, in $O(k \lg m)$ time, reporting the $k$
triples with maximum weight among all triples $(r,f,w)$ with $q \le r$.

An $\add(v, f, w, d)$ operation traverses the path $P_v = \langle x_1, \ldots,
x_h, v\rangle$ from the root of $T$ to $v$ (which is a leaf of $T$).  We insert
the triple $(d, f, w)$ into~$W_v$.  Each node $x_i$ in $P_v$ represents an edge
$(x,y)$ of $G$ and has two child edges $e_x$ and $e_y$ such that $x \in V_{e_x}$
and $y \in V_{e_y}$. Assume w.l.o.g.\ that $v \in V_{e_x}$, and  let $d' =
|\dist(\rho, y) - \dist(\rho, v)|$. We insert the triple $(d - d', f, w)$ into
the data structure $W_{e_y}$ associated with $e_y$. This is illustrated in
\cref{fig:centroid}.  This takes $O(\lg m)$ time per vertex in~$P_v$, $O(\lg n
\lg m)$ time in total.

A $\remove(v, f)$ operation traverses the path $P_v$ and deletes the triple
associated with $f$ from the data structure $W_e$ of every child edge $e$ of
every node $x_i$ in $P_v$, and from $W_v$.  By a similar analysis as for
$\add(v, f, w, d)$ operations, this takes $O(\lg n \lg m)$ time.

A $\sumq(v, d)$ query traverses the path $P_v$. For each edge $e$ on this path
with top endpoint $x_i$, $x_i$ represents an edge $(x, y) \in G$ such that
w.l.o.g.\ $e = e_y$. We query $W_e$ to report the total weight of all triples
$(r, f, w)$ in $W_e$ with $r \ge |\dist(\rho, y) - \dist(\rho, v)| - d$. We also
query $W_v$ to report the total weight of all facilities placed on $v$ itself.
This is illustrated in \cref{fig:centroid}.  We sum these weights retrieved from
$W_v$ and from all the edge data structures $W_e$ along $P_v$ and report the
resulting total.  Since we answer a $\sumq(v, d)$ query by querying $O(\lg n)$
data structures $W_e$ and~$W_v$, the cost is $O(\lg n \lg m)$.  A $\sumq(v)$
query is the same as a $\sumq(v, 0)$ query.

A $\topk(v, k, d)$ query traverses $P_v$.  For each edge $e$ on this path with
top endpoint $x_i$, $x_i$ represents an edge $(x, y) \in G$ such that w.l.o.g.\
$e = e_y$.  We query $W_e$ to retrieve the $k$ triples with maximum weight among
all triples $(r, f, w)$ in $W_e$ such that $r \ge |\dist(\rho, y) - \dist(\rho,
v)| - d$.  This takes $O(k \lg n \lg m)$ time for all edges on $P_v$.  We also
retrieve the $k$ facilities with maximum weight from $W_v$, which takes $O(k \lg
m)$ time.  The $k$ maximum-weight facilities affecting vertices at distance at
most $d$ from $v$ are among the $O(k \lg n)$ facilities retrieved by these
queries and can be found in $O(k \lg n)$ time using linear-time selection
\cite{blumTimeBoundsSelection1973}.  Thus, a $\topk(v, k, d)$ query takes $O(k
\lg n \lg m)$ time. A $\topk(v, k)$ query is the same as a $\topk(v, k, 0)$
query.

To prove the correctness of $\sumq(v, d)$ and $\topk(v, k, d)$ queries, note
that both queries query the same data structures $W_e$, with the same query
regions. A $\sumq$ query reports the total weight of all facilities in these
query regions. A $\topk$ query reports the $k$ maximum weight queries in these
regions. Both queries are correct if we can argue that if either query were to
\emph{report} all facilities in these query regions instead of summing their
weights or picking the $k$ facilities with maximum weight, then any facility
placed on some vertex $u$ is reported if and only if its effect radius $d'$
satisfies $d + d' \ge \dist(u, v)$, and any such facility is reported exactly
once.

So consider a facility $f$ with effect radius $d'$ placed on some vertex $u \in
G$, and let $v$ be another vertex $v \in G$. If $v = u$, then $f$ must be
reported because it affects $v$ no matter the query radius $d$. The facility $f$
does not belong to any data structure $W_e$ on the path $P_u = P_v$. Thus, if
$f$ is to be reported, it must be reported by the query $W_v$. Since placing $f$
on $u$ adds $f$ to $W_u = W_v$, a $\sumq(v,d)$ query reports the total weight of
all facilities in $W_v$, and a $\topk(v,k,d)$ query reports the $k$ facilities
with maximum weight in $W_v$, the corresponding reporting query would report all
facilities in $W_v$, including $f$.

If $v \ne u$, then let $x_i$ be the highest vertex on the path from $u$ to $v$
in $T$.  This vertex represents an edge $(x,y)$ such that w.l.o.g.\ $u \in
V_{e_x}$ and $v \in V_{e_y}$. In this case, $f$ is not stored in $W_v$, and
$e_y$ is the only edge on the path $P_v$ that is a pendant edge of $P_u$.  Thus,
$W_{e_y}$ is the only data structure considered by a $\sumq(v,d)$ or
$\topk(v,k,d)$ query that stores $f$. In $W_{e_y}$, $f$ is stored with radius $r
= d' - |\dist(\rho,y) - \dist(\rho,u)| = d' - \dist(u,y)$. The path from $u$ to
$v$ in $G$ passes through $y$, so $\dist(u,v) = \dist(u,y) + \dist(y,v)$.
Therefore, $\dist(u,v) \le d + d'$ if and only if $q = |\dist(\rho,v) -
\dist(\rho,y)| - d = \dist(v,y) - d \le d' - \dist(u,y) = r$.  The reporting
version of a $\sumq(v,d)$ or $\topk(v,k,d)$ query reports all triples $(r,f,w)$
in $W_{e_y}$ with $r \ge q$.  Thus, $f$ is reported if and only if $d + d' \ge
\dist(u,v)$.

To bound the size of the data structure, note that $T$ has size $O(n)$, each
data structure $W_x$, with $x$ a leaf or edge of $T$, has size linear in the
number of triples it stores, and each facility placed on some vertex $v$ is
stored in $W_v$ and in the data structure $W_e$ associated with each edge $e$ on
the path $P_v$. Since this path has length $O(\lg n)$, each facility is stored
$O(\lg n)$ times.  Thus, the SOLE data structure for trees of degree at most 3
has size $O(n + m \lg n)$.

To obtain a SOLE data structure for arbitrary trees (of degree greater than~3),
we can transform any such tree $G$ into a tree $G'$ whose nodes have degree at
most $3$ by replacing every high-degree vertex $u$ in $G$ with a degree-3
subtree $G'_u$ whose edges all have length 0 (see
Figs.~\labelcref{fig:degree-3},\subref{fig:centroid}).  We choose an arbitrary
vertex in $G_u'$ as the representative of $u$ in $G'$. This ensures that the
distances between vertices in $G$ and between their representatives in $G'$ are
the same.  Thus, we can support $\add$, $\remove$, $\sumq$, and $\topk$ queries
on $G$ by building a SOLE data structure on $G'$ instead.  This finishes the
proof of \cref{thm:trees}.

\section{A SOLE Data Structures for Separable Graphs}

\label{sec:sp_graphs}

In this section, we generalize the result from \cref{sec:trees} to $t$-separable
graphs.  We call a graph $G$ \emph{$t$-separable}, for some constant $t$, if it
has a \emph{$t$-separator decomposition $C$} of the following structure (see
Figs.~\labelcref{fig:graph},\subref{fig:graph-decomposition}):
\begin{itemize}
  \item $C$ is an unrooted tree with $O(n)$ nodes, all of which have degree at
  most 3.
  \item Each edge $e$ of $C$ has an associated subset $S_e \subseteq V$ of
  vertices of $G$ of size $|S_e| \le t$. We call $S_e$ the \emph{(edge) bag}
  associated with $e$.
  \item Every vertex of $G$ belongs to at least one bag of $C$.
  \item Let $C_1$ and $C_2$ be the subtrees of $C$ obtained by removing any edge
  $e$ from~$C$, and let $V_i$, $i \in \{1, 2\}$, be the union of the bags of all
  edges in $C_i$. Then any path from a vertex in $V_1$ to a vertex in $V_2$
  includes at least one vertex in $S_e$. In other words, $S_e$ separates the
  vertices in $V_1$ from the vertices in $V_2$.
\end{itemize}
This definition of a $t$-separator decomposition is similar to both a tree
decomposition \cite{robertsonGraphMinorsObstructions1991} and a branch
decomposition \cite{robertsonGraphMinorsObstructions1991}, but the properties of
a $t$-separator decomposition are weaker than those of both a tree-decomposition
and a branch decomposition.  In particular, a branch decomposition of width $b$
and degree 3 is easily seen to be a $b$-separator decomposition, and a nice tree
decomposition~$C$~\cite{bodlaenderEfficientConstructiveAlgorithms1996} of width
$w$ gives rise to a $(w+1)$-separator decomposition by defining the (edge) bag
associated with each edge $(v,w) \in C$ to be the union of the (node) bags
associated with $v$ and $w$.  However, a $t$-separator decomposition does not
require a bijection between the edges of $G$ and the leaves of $C$, as required
by a branch decomposition. A tree decomposition requires that for every edge
$(v,w)$ of $G$, there exists a (node) bag that contains both $v$ and $w$, a
condition not imposed by a $t$-separator decomposition.  Thus, every graph of
branchwidth $b$ has a $t$-separator decomposition with $t \le b$, every graph of
treewidth $w$ has a $t$-separator decomposition with $t \le w + 1$, but there
may exist graphs for which these inequalities are strict.

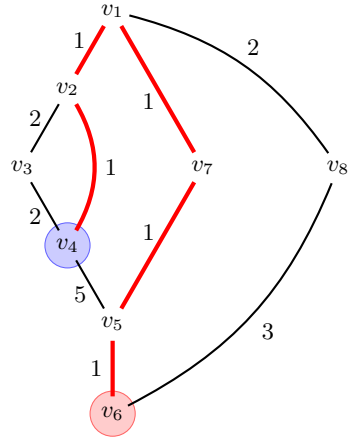
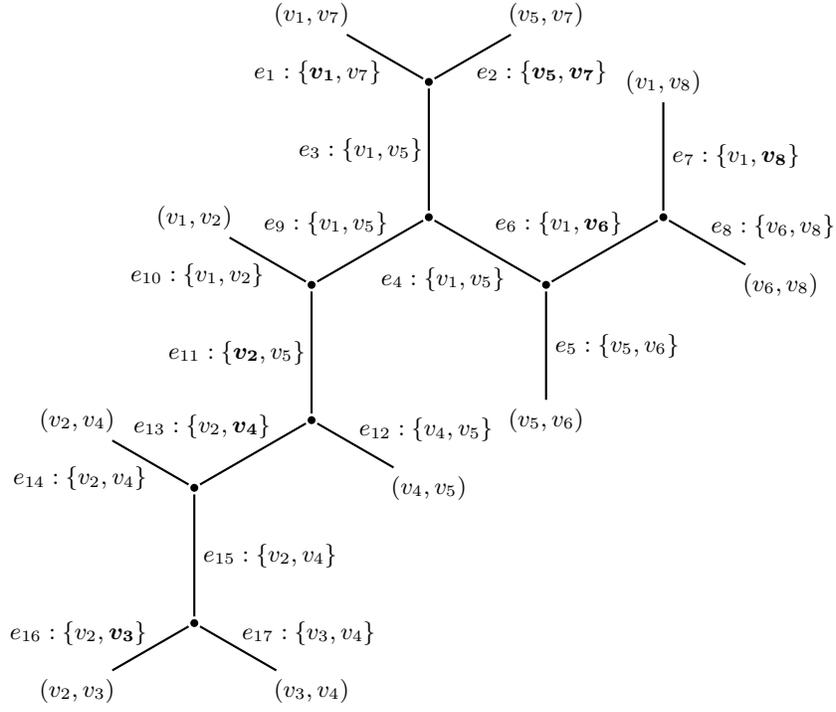
\begin{figure}[p]
  \centering

  \subfloat[\label{fig:graph}]{%
    \begin{tikzpicture}[
      vertex/.style={circle,inner sep=1pt},
      edge/.style={thick}
    ]
     \path node [vertex] (v1) {$v_1$}
     ++(240:1.2) node [vertex] (v2) {$v_2$}
     ++(240:1.2) node [vertex] (v3) {$v_3$}
     ++(300:1.2) node [vertex] (v4) {$v_4$}
     ++(300:1.2) node [vertex] (v5) {$v_5$}
     +(270:1.2) node [vertex] (v6) {$v_6$}
     (v3) +(0:2.4) node [vertex] (v7) {$v_7$}
     (v7) +(0:1.8) node [vertex] (v8) {$v_8$};
     \draw [edge]
     (v2) to node [left,pos=0.3]                    {$2$} (v3)
     (v3) to node [left,pos=0.7]                    {$2$} (v4)
     (v4) to node [left,pos=0.7]                    {$5$} (v5)
     (v1) to [bend left=20] node [above,xshift=4pt] {$2$} (v8)
     (v8) to [bend left=20] node [below,xshift=6pt] {$3$} (v6);
     \path [ultra thick,draw=red]
     (v1) to node [left,pos=0.3]                    {$1$} (v2)
     (v1) to node [left,pos=0.6]                    {$1$} (v7)
     (v7) to node [left,pos=0.4]                    {$1$} (v5)
     (v5) to node [left]                            {$1$} (v6)
     (v2) to [bend left] node [right]               {$1$} (v4);
     \begin{scope}[on background layer]
       \node [draw=red!60,fill=red!20,circle,inner sep=0pt,minimum size=6mm] at (v6) {};
       \node [draw=blue!60,fill=blue!20,circle,inner sep=0pt,minimum size=6mm] at (v4) {};
      \end{scope}
    \end{tikzpicture}
  }

  \bigskip
  \bigskip

  \subfloat[\label{fig:graph-decomposition}]{%
    \begin{tikzpicture}[
      vertex/.style={rectangle,inner sep=3pt},
      inner node/.style={fill,circle,inner sep=0pt,minimum size=3pt,outer sep=1pt},
      edge/.style={thick},
      edge label/.style={inner sep=1pt}
    ]
      \path node [inner node] (x1) {}
      ++ (90:1.8) node [inner node] (x2) {}
      +(150:1.8) node [vertex] (e7) {$(v_1,v_7)$}
      +(30:1.8) node [vertex] (e8) {$(v_5, v_7)$}
      (x1) ++(330:1.8) node [inner node] (x3) {}
      +(270:1.8) node [vertex] (e6) {$(v_5, v_6)$}
      ++(30:1.8) node [inner node] (x4) {}
      +(90:1.8) node [vertex] (e9) {$(v_1, v_8)$}
      +(330:1.8) node [vertex] (e10) {$(v_6, v_8)$}
      (x1) ++(210:1.8) node [inner node] (x5) {}
      +(150:1.8) node [vertex] (e1) {$(v_1, v_2)$}
      ++(270:1.8) node [inner node] (x6) {}
      +(330:1.8) node [vertex] (e5) {$(v_4, v_5)$}
      ++(210:1.8) node [inner node] (x7) {}
      +(150:1.8) node [vertex] (e4) {$(v_2, v_4)$}
      ++(270:1.8) node [inner node] (x8) {}
      +(210:1.8) node [vertex] (e2) {$(v_2, v_3)$}
      +(330:1.8) node [vertex] (e3) {$(v_3, v_4)$};
      \draw [edge]
      (e7) to node [edge label,anchor=north east]         {$e_1: \{\bm{v_1},v_7\}$} (x2)
      (e8) to node [edge label,anchor=north west]         {$e_2: \{\bm{v_5},\bm{v_7}\}$} (x2)
      (x2) to node [edge label,anchor=east,xshift=-1pt]   {$e_3: \{v_1,v_5\}$} (x1)
      (x1) to node [edge label,anchor=north east,pos=0.7] {$e_4: \{v_1,v_5\}$} (x3)
      (x3) to node [edge label,anchor=west,xshift=2pt]    {$e_5: \{v_5,v_6\}$} (e6)
      (x3) to node [edge label,anchor=south east,pos=0.7] {$e_6: \{v_1,\bm{v_6}\}$} (x4)
      (x4) to node [edge label,anchor=west,xshift=2pt]    {$e_7: \{v_1,\bm{v_8}\}$} (e9)
      (x4) to node [edge label,anchor=south west]         {$e_8: \{v_6,v_8\}$} (e10)
      (x1) to node [edge label,anchor=south east,pos=0.3] {$e_9: \{v_1,v_5\}$} (x5)
      (x5) to node [edge label,anchor=north east]         {$e_{10}: \{v_1,v_2\}$} (e1)
      (x5) to node [edge label,anchor=east,xshift=-1pt]   {$e_{11}: \{\bm{v_2},v_5\}$} (x6)
      (x6) to node [edge label,anchor=south west]         {$e_{12}: \{v_4,v_5\}$} (e5)
      (x6) to node [edge label,anchor=south east,pos=0.3] {$e_{13}: \{v_2,\bm{v_4}\}$} (x7)
      (x7) to node [edge label,anchor=north east]         {$e_{14}: \{v_2,v_4\}$} (e4)
      (x7) to node [edge label,anchor=west,xshift=2pt]    {$e_{15}: \{v_2,v_4\}$} (x8)
      (x8) to node [edge label,anchor=south east]         {$e_{16}: \{v_2,\bm{v_3}\}$} (e2)
      (x8) to node [edge label,anchor=south west]         {$e_{17}: \{v_3,v_4\}$} (e3)
      ;
    \end{tikzpicture}
  }

  \caption{Caption on next page.}
\end{figure}
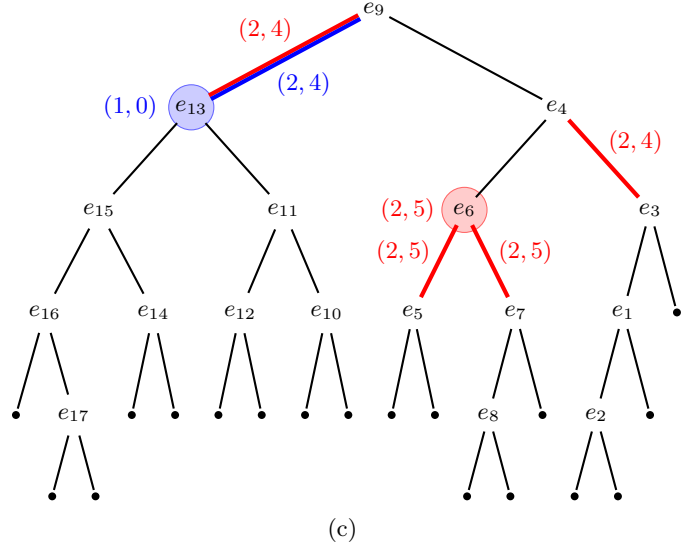
\begin{figure}[t]
  \centering\ContinuedFloat
  \subfloat[\label{fig:graph-centroid}]{%
    \begin{tikzpicture}[
      vertex/.style={circle,inner sep=1pt},
      leaf/.style={fill,circle,inner sep=0pt,minimum size=3pt,outer sep=1pt},
      edge/.style={thick}
    ]
    \path
    node [leaf] (l1) {}
    node [leaf,right=4mm of l1] (l2) {}
    (barycentric cs:l1=0.5,l2=0.5) node [vertex,anchor=south,yshift=8mm] (e17) {$e_{17}$}
    node [leaf,left=4mm of e17] (l3) {}
    node [leaf,right=4mm of e17] (l4) {}
    node [leaf,right=4mm of l4] (l5) {}
    node [leaf,right=4mm of l5] (l6) {}
    node [leaf,right=4mm of l6] (l7) {}
    node [leaf,right=4mm of l7] (l8) {}
    node [leaf,right=4mm of l8] (l9) {}
    node [leaf,right=4mm of l9] (l16) {}
    node [leaf,right=4mm of l16] (l17) {}
    node [vertex,right=4mm of l17] (e8) {$e_8$}
    node [leaf,right=4mm of e8] (l15) {}
    node [vertex,right=4mm of l15] (e2) {$e_2$}
    node [leaf,right=4mm of e2] (l14) {}
    node [leaf,anchor=east,xshift=-2mm] at (e8 |- l1) (l10) {}
    node [leaf,anchor=west,xshift=2mm] at (e8 |- l1) (l11) {}
    node [leaf,anchor=east,xshift=-2mm] at (e2 |- l1) (l12) {}
    node [leaf,anchor=west,xshift=2mm] at (e2 |- l1) (l13) {}
    ({barycentric cs:l3=0.5,e17=0.5} |- e17.north) node [vertex,anchor=south,yshift=8mm] (e16) {$e_{16}$}
    ({barycentric cs:l4=0.5,l5=0.5} |- e16) node [vertex] (e14) {$e_{14}$}
    ({barycentric cs:l6=0.5,l7=0.5} |- e16) node [vertex] (e12) {$e_{12}$}
    ({barycentric cs:l8=0.5,l9=0.5} |- e16) node [vertex] (e10) {$e_{10}$}
    ({barycentric cs:l16=0.5,l17=0.5} |- e16) node [vertex] (e5) {$e_5$}
    ({barycentric cs:e8=0.5,l15=0.5} |- e16) node [vertex] (e7) {$e_7$}
    ({barycentric cs:e2=0.5,l14=0.5} |- e16) node [vertex] (e1) {$e_1$}
    node [leaf,right=4mm of e1] (l18) {}
    ({barycentric cs:e16=0.5,e14=0.5} |- e16.north) node [vertex,anchor=south,yshift=8mm] (e15) {$e_{15}$}
    ({barycentric cs:e12=0.5,e10=0.5} |- e15) node [vertex] (e11) {$e_{11}$}
    ({barycentric cs:e5=0.5,e7=0.5} |- e15) node [vertex] (e6) {$e_6$}
    ({barycentric cs:e1=0.5,l18=0.5} |- e15) node [vertex] (e3) {$e_3$}
    ({barycentric cs:e15=0.5,e11=0.5} |- e15.north) node [vertex,anchor=south,yshift=8mm] (e13) {$e_{13}$}
    ({barycentric cs:e6=0.5,e3=0.5} |- e13) node [vertex] (e4) {$e_4$}
    ({barycentric cs:e13=0.5,e4=0.5} |- e13.north) node [vertex,anchor=south,yshift=8mm] (e9) {$e_9$};
    \draw [edge] (l1) -- (e17) -- (l2)
    (l3) -- (e16) -- (e17)
    (l4) -- (e14) -- (l5)
    (l6) -- (e12) -- (l7)
    (l8) -- (e10) -- (l9)
    (l16) -- (e5) -- (l17)
    (l10) -- (e8) -- (l11)
    (l12) -- (e2) -- (l13)
    (e8) -- (e7) -- (l15)
    (e2) -- (e1) -- (l14)
    (e1) -- (e3) -- (l18)
    (e16) -- (e15) -- (e14)
    (e12) -- (e11) -- (e10)
    (e15) -- (e13) -- (e11)
    (e6) -- (e4)
    (e9) -- (e4);
    \draw [red,ultra thick]
    (e5) to node [inner sep=1pt,anchor=south east,pos=0.4] {$(2,5)$} (e6)
    (e6) to node [inner sep=1pt,anchor=south west,pos=0.6] {$(2,5)$} (e7)
    (e4) to node [inner sep=1pt,anchor=south west] {$(2,4)$} (e3);
    \begin{scope}
      \path (e9.center) +(180:70pt) coordinate (start);
      \clip (start) -- (e9.center) -- (e13.center) -- cycle;
      \draw [line width=3.2pt,red] (e13) to node [inner sep=1pt,anchor=south east,pos=0.6] {$(2,4)$} (e9);
    \end{scope}
    \begin{scope}
      \path (e13.center) +(0:70pt) coordinate (start);
      \clip (start) -- (e9.center) -- (e13.center) -- cycle;
      \draw [line width=3.2pt,blue] (e13) to node [inner sep=1pt,anchor=north west,pos=0.4] {$(2,4)$} (e9);
    \end{scope}
    \begin{scope}[on background layer]
      \node [draw=red!60,fill=red!20,circle,inner sep=0pt,minimum size=6mm] at (e6) {};
      \node [draw=blue!60,fill=blue!20,circle,inner sep=0pt,minimum size=6mm] at (e13) {};
    \end{scope}
    \node [red,left=0.5mm of e6] {$(2, 5)$}; 
    \node [blue,left=0.5mm of e13] {$(1, 0)$}; 
    \end{tikzpicture}
  }

  \captionsetup{format=cont}
  \caption{A series-parallel graph $G$ (\subref{fig:graph}), a branch
  decomposition $C$ of $G$ of width~$2$ (\subref{fig:graph-decomposition}),
  which is also a $2$-separator decomposition of $G$, and the centroid
  decomposition $T$ of $C$ (\subref{fig:graph-centroid}). Each edge in
  (\subref{fig:graph}) is labelled with its length. Each edge in
  (\subref{fig:graph-decomposition}) is labelled with its name and its
  corresponding bag $S_e$. A facility with effect radius 5 placed on vertex
  $v_6$ affects vertex $v_4$, since the path shown in red in
  (\subref{fig:graph}) has length 5. If we assume that $e_{v_6} = e_6$ and
  $e_{v_4} = e_{13}$, as indicated by the bold vertices in
  (\subref{fig:graph-decomposition}), then $f$ is stored in the node data
  structure of $e_6$ and in the edge data structures of the red edges in
  (\subref{fig:graph-centroid}), with the pairs of radii shown in red. A
  $\sumq(v_4)$ or $\topk(v_4,k)$ query queries the vertex data structure of
  $e_{13}$ and the edge data structure of the blue edge in
  (\subref{fig:graph-centroid}), with the pairs of radii shown in blue. The
  facility $f$ is added to the query result because $2 \le 2$ and $4 \le 4$. One
  of these two conditions would have sufficed.}
\end{figure}

In this section, we prove that

\begin{theorem}
  \label{thm:separable}
  If $G$ is a $t$-separable graph on $n$ vertices, for some constant $t$, then
  there is a SOLE data structure for it supporting $\add$, $\remove$, and
  $\sumq$ operations in $O (\lg n \lg^t m)$ time, and $\topk(v, k, d)$
  operations in $O(k \lg n \lg^t m)$ time. The size of this data structure is
  $O(tn \lg n + m \lg^{t-1} m \lg n)$.  The costs of $\add$ and $\remove$
  operations are amortized.
\end{theorem}

Let $C$ be a $t$-separator decomposition for $G$.  Since $C$ is an unrooted tree
whose nodes have degree at most $3$, we can once again construct its centroid
decomposition $T$ (see \cref{fig:graph-centroid}). Each leaf of $T$ corresponds
to a node of $C$, and each internal node of $T$ corresponds to an edge $e$ of
$C$.  Since $C$ has size $O(n)$, the height of $T$ is $O(\lg n)$.  Since a
vertex $v$ of $G$ may be contained in more than one bag of $C$, we choose an
arbitrary bag $S_e$ of $C$ that contains $v$ and refer to $e$ as $e_v$, as
indicated by the bold vertex labels in \cref{fig:graph-decomposition}.

We obtain a SOLE data structure for $G$ by augmenting each internal node $e$ of
$T$ with two data structures $W_e$ and $D_e$ and augmenting each edge $a$ of $T$
with a data structure $W_a$. We refer to $W_e$ as a \emph{node data structure}
and to $W_a$ as an \emph{edge data structure}. Each data structure $W_x$, with
$x$ a node or an edge of~$T$, stores a number of facilities $f$ as tuples $(r_1,
\ldots, r_{t'}, f, w)$. If $x = e$ is a node of $T$ or $x = a$ is an edge of $T$
with top endpoint $e$, then $t' = |S_e| \le t$.  Again, $W_x$ consists of two
trees $R_x$ and $F_x$ over the set of tuples stored in $W_x$.  $F_x$~stores
these tuples as a binary search tree with the facility $f$ as the key for each
tuple $(r_1, \ldots, r_{t'}, f, w)$ in $W_x$.  $R_x$ is a ``$t'$-dimensional
range sum priority search tree'' (see \cref{sec:combinations}) over the points
defined by the coordinates $(r_1, \ldots, r_{t'})$. This is a $t'$-dimensional
range tree
\cite{bentleyDecomposableSearchingProblems1979,willardAddingRangeRestriction1985}
augmented to support $t'$-dimensional range sum queries in $O\bigl(\lg^{t'}
m\bigr)$ time and $t'$-dimensional range top-$k$ queries in $O\bigl(k \lg^{t'}
m\bigr)$ time.

The data structure $D_e$ associated with each internal node $e$ of $T$ stores
the distances from every vertex $v$ such that $e_v$ is a descendant of $e$ in
$T$ to all vertices in $S_e$.

As discussed in \cref{sec:combinations}, a $t'$-dimensional range sum priority
search tree supports insertions and deletions in $O\bigl(\lg^{t'} m\bigr)$
amortized time.  Thus, each data structure $W_x$ supports insertion of a new
tuple $(r_1, \ldots, r_{t'}, f, w)$ and the deletion of the tuple associated
with a facility $f$ in $O\bigl(\lg^{t'} m\bigr) = O(\lg^t m)$ amortized time.

To support $\sumq$ and $\topk$ queries, we need to support range sum queries and
range top-$k$ queries on $W_x$, but with an uncommon query range.  A query point
$q = (q_1, \ldots, q_{t'})$ defines a query range $\RR^{t'} \setminus ((-\infty,
q_1) \times (-\infty, q_2) \times \cdots \times (-\infty, q_{t'}))$, that is,
the complement of $t'$-sided range query. We need to support range sum queries
and range top-$k$ queries for any such complement of a $t'$-sided range query.

An easy way to support such a query is to decompose it into $t'$ ``normal''
range queries, with query ranges
\begin{itemize}
\item $[q_1,\infty) \times (-\infty, \infty) \times \cdots \times (-\infty, \infty)$,
\item $(-\infty, q_1) \times [q_2,\infty) \times (-\infty, \infty) \times \cdots \times (-\infty, \infty)$,
\item $(-\infty, q_1) \times (-\infty,q_2) \times [q_3,\infty) \times (-\infty, \infty) \times \cdots \times (-\infty, \infty)$,
\item $\cdots$
\item $(-\infty, q_1) \times (-\infty,q_2) \times \cdots \times (-\infty,q_{t'-1}) \times [q_{t'},\infty)$.
\end{itemize}
This allows us to support range sum and range top-$k$ queries in $O\bigl(t'
\lg^{t'} m\bigr)$ and $O\bigl(t' k \lg^{t'} m\bigr)$ time, respectively, which
is $O\bigl(\lg^{t'} m\bigr)$ and $O\bigl(k \lg^{t'} m\bigr)$ time because $t'
\le t$ and $t$ is a constant.

A better way to support range queries with query ranges of the form $\RR^{t'}
\setminus ((-\infty, q_1) \times (-\infty, q_2) \times \cdots \times (-\infty,
q_{t'}))$ without the factor $t'$ overhead is to implement them directly on the
$t'$-dimensional range sum priority search tree. To answer a range sum query
with such a query range, we answer a $1$-dimensional range sum query with query
range $[q_1,\infty)$ on the level-$1$ tree of $R_x$. This query traverses the
path corresponding to $q_1$ in the level-$1$ tree of $R_x$.  For the root of
each subtree to the left of this path, we answer a $(t'-1)$-dimensional range
sum query with query range $\RR^{t'-1} \setminus ((-\infty, q_2) \times
(-\infty, q_3) \times \cdots \times (-\infty, q_{t'}))$ on the
$(t'-1)$-dimensional range sum priority search tree associated with this root.
The final result is the sum of the totals produced by these queries, including
the $1$-dimensional range sum query on the level-$1$ tree of $R_x$.  Thus, a
range sum query with the complement of a $t'$-sided range query as the query
range has the same cost as a ``normal'' orthogonal range sum query, $O\bigl(\lg^{t'}
m\bigr)$.

Similarly, to support a $t'$-dimensional range top-$k$ query with query range $Q
= \RR^{t'} \setminus ((-\infty, q_1) \times (-\infty, q_2) \times \cdots \times
(-\infty, q_{t'}))$, we answer a $1$-dimensional range top-$k$ query on the
level-$1$ tree of $R_x$, with query range $[q_1, \infty)$. This query traverses
the path corresponding to $q_1$ in the level-$1$ tree of $R_x$.  For the root of
each subtree to the left of this path, we answer a $(t'-1)$-dimensional range
top-$k$ query with query range $\RR^{t'-1} \setminus ((-\infty, q_2) \times
(-\infty, q_3) \times \cdots \times (-\infty, q_{t'}))$ on the
$(t'-1)$-dimensional range sum priority search tree associated with this root.
The top $k$ tuples in the query range $Q$ are easily seen to be among the $O(k
\log n)$ elements reported by these $(t'-1)$-dimensional range top-$k$ queries
and by the $1$-dimensional range top-$k$ query on the level-$1$ tree.  The top
$k$ tuples can now be found in $O(k \lg n)$ time using linear-time selection
\cite{blumTimeBoundsSelection1973}. Thus, a range top-$k$ query with the
complement of a $t'$-sided range query as the query range takes $O\bigl(k
\lg^{t'} m\bigr)$ time, just as a ``normal'' orthogonal range top-$k$ query
does.

We are ready to discuss how to support $\add$, $\remove$, $\sumq$, and $\topk$
operations on our SOLE data structure for $t$-separable graphs.

An $\add(v,f,w,d)$ operation traverses the path $P_v = \langle e_1, \ldots, e_h
= e_v\rangle$ in $T$ from the root of $T$ to the node $e_v$. For each node $e_i$
in $P_v$, let $S_{e_i} = \{v_1, \ldots, v_{t'}\}$, and let $d_1, \ldots, d_{t'}$
be the distances from $v$ to $v_1, \ldots, v_{t'}$. Then we add the tuple $(r_1,
\ldots, r_{t'}, f, w)$ to $W_{e_i}$, where $r_j = d - \dist(v, v_j)$ for all $1
\le j \le t'$. If $i = h$, we also add the tuple $(r_1, \ldots, r_{t'}, f, w)$
to the edge data structures $W_a$ associated with the child edges of $e_i$ in
$T$. If $i < h$, we add $(r_1, \ldots, r_{t'}, f, w)$ only to the data structure
$W_a$ of the child edge $a$ of $e_i$ that does \emph{not} belong to the path
$P_v$.  This is illustrated in \cref{fig:graph-centroid}.  Since each data
structure $W_x$ supports insertions in amortized $O(\lg^t m)$ time, the
amortized cost of an $\add$ operation is thus $O(\lg n \lg^t m)$.

A $\remove(v, f)$ operation traverses the path $P_v = \langle e_1, \ldots, e_k =
e_v\rangle$ and removes $f$ from each data structure $W_{e_i}$ and from the data
structures $W_a$ associated with both child edges of each node $e_i$ in $P_v$.
Since each data structure $W_x$ supports deletions in amortized $O(\lg^t m)$
time, the amortized cost of a $\remove$ operation is thus $O(\lg n \lg^t m)$.

A $\sumq(v, d)$ query traverses the path $P_v = \langle e_1, \ldots, e_h =
e_v\rangle$. For each node $e_i$ in $P_v$, let $S_{e_i} = \{v_1, \ldots,
v_{t'}\}$, and let $Q = \RR^{t'} \setminus ((-\infty, q_1) \times (-\infty, q_2)
\times \cdots \times (-\infty, q_{t'}))$, where $q_j = \dist(v, v_i) - d$, for
all $1 \le j \le t'$.  If $1 \le i < h$, then we answer a range sum query with
query range $Q$ on the edge data structure~$W_a$, where $a = (e_i, e_{i+1})$ is
the child edge of $e_i$ that belongs to $P_v$. If $i = h$, then we answer a
range sum query with query range $Q$ on $W_{e_i}$.  This is illustrated in
\cref{fig:graph-centroid}.  The result of the $\sumq(v, d)$ query is the sum of
the results reported by all these range sum queries.  Since a range sum query on
each data structure $W_x$ can be answered in $O\bigl(\lg^{t'} m\bigr) = O(\lg^t
m)$ time, the cost of a $\sumq(v, d)$ query is thus $O(\lg n \lg^t m)$. A
$\sumq(v)$ query is the same as a $\sumq(v, 0)$ query.

A $\topk(v, k, d)$ query traverses the path $P_v = \langle e_1, \ldots, e_h =
e_v\rangle$. For each node $e_i$ in $P_v$,  let $S_{e_i} = \{v_1, \ldots,
v_{t'}\}$, and let $Q = \RR^{t'} \setminus ((-\infty, q_1) \times (-\infty, q_2)
\times \cdots \times (-\infty, q_{t'}))$, where $q_j = \dist(v, v_i) - d$, for
all $1 \le j \le t'$.  If $1 \le i < h$, then we ask a range top-$k$ query with
query range $Q$ on the edge data structure~$W_a$, where $a = (e_i, e_{i+1})$ is
the child edge of $e_i$ that belongs to $P_v$. If $i = h$, then we answer a
range top-$k$ query with query range $Q$ on $W_{e_i}$. The result of the
$\topk(v, k, d)$ query is the list of the $k$ maximum-weight facilities among
the $O(k \lg n)$ facilities reported by all these range top-$k$ queries. These
$k$ facilities can be found in $O(k \lg n)$ time using linear-time selection
\cite{blumTimeBoundsSelection1973}.  Each query on a data structure $W_x$ takes
$O(k \lg^t m)$ time. Thus, the total cost of a $\topk(v, k, d)$ query is $O(k
\lg n \lg^t m)$. A $\topk(v, k)$ query is the same as a $\topk(v, k, 0)$ query.

To establish the correctness of $\sumq(v, d)$ and $\topk(v, k, d)$ queries,
observe that, similar to \cref{sec:trees}, both queries query the same node and
edge data structures, with the same query ranges. Thus, it suffices to prove
that if either query reported all facilities in these query ranges, it would
report any facility with effect radius $d'$ placed on some vertex $u$ if and
only if $d + d' \ge \dist(u, v)$, and each such facility is reported exactly
once.

So let $f$ be a facility with effect radius $d'$ placed on some vertex $u \in
G$, and let $v$ by any other vertex $v \in G$. Let $e$ be the lowest common
ancestor (LCA) of $e_u$ and $e_v$ in $T$, and let $S_e = \{v_1, \ldots,
v_{t'}\}$ We distinguish two cases:

If $e_v$ is a proper descendant of $e$, then $f$ is not stored in $W_{e_v}$ and
the only edge data structure in $P_v$ that stores $f$ is the data structure
$W_a$ corresponding to the child edge $a$ of $e$ on the path from $e$ to $e_v$.
Thus, $f$ is reported at most once by the reporting version of a $\sumq(v, d)$
or $\topk(v, k, d)$ query. This query queries $W_a$ with query region $Q =
\RR^{t'} \setminus ((-\infty, q_1) \times \cdots (-\infty, q_{t'}))$, where $q_j
= \dist(v_j) - d$ for all $1 \le j \le t'$. Since $e$ is the LCA of $e_u$ and
$e_v$ in $T$, any path from $u$ to $v$ in $G$ must include at least one vertex
in $S_e$. Assume w.l.o.g.\ that $v_1$ is one such vertex.  Then $\dist(u,v) =
\dist(v,v_1) + \dist(u, v_1)$ and $\dist(u, v) \le \dist(v, v_j) + \dist(u,
v_j)$ for all $1 < j \le t'$.  The facility $f$ is stored in $W_a$ as the tuple
$(r_1, \ldots, r_{t'}, f, w)$ with $r_j = d' - \dist(u, v_j)$ for all $1 \le j
\le t'$. Thus, $(r_1, \ldots, r_{t'}) \in Q$ if and only if there exists an
index $1 \le j \le t'$ such that $d' - \dist(u, v_j) \ge \dist(v, v_j) - d$,
that is, $d + d' \ge \dist(u, v_j) + \dist(v, v_j)$. Since $\dist(u, v_1) +
\dist(v, v_1) = \dist(u, v)$ and $\dist(u, v_j) + \dist(v, v_j) \ge \dist(u,v)$
for all $1 \le j \le t'$, this is true if and only if $d + d' \ge \dist(u, v)$.
Thus, $f$ is reported if and only if $d + d' \ge \dist(u, v)$.

If $e_v$ is not a proper descendant of $e$, then $e_v = e$ and $P_v \subseteq
P_u$. Thus, $f$~is not stored in any edge data structure along $P_v$, but it is
stored in $W_e = W_{e_v}$, as the tuple $(r_1, \ldots, r_{t'}, f, w)$ with $r_j
= d' - \dist(u, v_j)$ for all $1 \le j \le \nobreak t'$.  Thus, $f$ is reported
at most once by the reporting version of a $\sumq(v, d)$ or $\topk(v, k, d)$
query. This query queries $W_{e_v}$ with query region $Q = \RR^{t'} \setminus
((-\infty, q_1) \times \cdots (-\infty, q_{t'}))$, where $q_j = \dist(v_j) - d$
for all $1 \le j \le t'$. Since $v \in S_{e_v}$, we can assume w.l.o.g.\ that $v
= v_1$.  Then $\dist(u,v) = \dist(v,v_1) + \dist(u, v_1)$ and $\dist(u, v) \le
\dist(v, v_j) + \dist(u, v_j)$ for all $1 < j \le t'$. The same analysis as in
the previous case now shows that $f$ is reported by the query on $W_{e_v}$ if
and only if $d + d' \ge \dist(u, v)$.

It remains to bound the size of the SOLE data structure.  Once again, the tree
$T$ has size $O(n)$ and height $O(\lg n)$. For every vertex $v \in G$, the
distance data structure $D_e$ of every ancestor node $e$ of $e_v$ in $T$ stores
the distances from $v$ to all $t' \le t$ vertices in $S_e$. Thus, each vertex
$v$ contributes at most $t$ to the size of each of $O(\lg n)$ distance data
structures.  The total size of the distance data structures is thus $O(tn \lg
n)$. Each facility $f$ placed on some vertex $u$ is stored in the node data
structures of the $O(\lg n)$ nodes along the path $P_u$ and in one or two edge
data structures of child edges of these nodes. Thus, every facility is stored in
$O(\lg n)$ data structures $W_x$. Each such data structure is a $t'$-dimensional
range sum priority search tree, where $t' \le t$. Thus, if it stores $s$
facilities, it has size $O(s \lg^{t-1} s)$. The total size of all node and edge
data structures is thus $O(m \lg^{t-1} m \lg n)$. This finishes the proof of
\cref{thm:separable}.

\appendix

\section{Semigroups}

\label{sec:semigroup}

A semigroup is a set $S$ equipped with an addition operation $+ : S \times S
\rightarrow S$ that is associative: $x + (y + z) = (x + y) + z$ for all $x, y, z
\in S$. What is \emph{not} required is that there must exist a \emph{unit
element} $0 \in S$ such that $0 + x = x = x + 0$, nor does every element $x \in
S$ need to have an \emph{inverse} $-x \in S$ such that $x + (-x) = 0$.

Examples of semigroups include standard number types with addition or
multiplication, any totally ordered set $S$ with $x + y = \min(x, y)$ or $x + y
= \max(x, y)$, a set of sequences with $x + y$ the concatenation of $x$ and $y$,
and many more.

\section{Standard Data Structures}

We need a number of standard data structures as building blocks for our SOLE
data structures. We review them here. All the binary search trees on which these
data structures are based are ``leaf-oriented'', that is, the data items they
store are stored at (or at least associated with) the leaves of the tree;
internal nodes only store search information to find the leaf where each item is
stored.

\subsection{Priority Search Tree}

\label{sec:priority-search-tree}

A priority search tree \cite{mccreightPrioritySearchTrees1985} over a set of
points $P$ in 2-d is a binary search tree $T$ storing the points in $P$ by their
$x$-coordinates. There is one leaf associated with every point in $P$. However,
points aren't necessarily stored at the leaves with which they are associated;
every point is stored at some ancestor of its corresponding leaf, defined as
follows: The point with maximum $y$-coordinate is stored at the root of $T$.
The left child of the root stores the point with maximum $y$-coordinate among
the points associated with the leaves in the left subtree, excluding the point
already stored at the root if this point came from the left subtree. The point
stored at the right child of the root is defined analogously, as are the points
stored at all other nodes in the tree.

A priority search tree can be built in linear time if the points are given
sorted by their $x$-coordinates. Insertions and deletions take worst-case
$O(\lg n)$ if the underlying binary search tree structure is a red-black
tree.

The query operation on priority search trees that we are interested in is
\emph{range top-$k$ queries}: Given a range of $x$-coordinates $[x_1, x_2]$,
report the $k$ points with maximum $y$-coordinates among the points with
$x$-coordinates in the interval $[x_1, x_2]$. Such an operation can be supported
in $O(\lg n + k(\lg\lg n + \lg k)) = O(k \lg n)$ time: Traverse the two paths
$P_1$ and $P_2$ in $T$ corresponding to the coordinates $x_1$ and $x_2$. Both
paths have length $O(\lg n)$. The part of $T$ between these two paths decomposes
into $O(\lg n)$ subtrees $T_1, \ldots, T_h$, at most one subtree per node in
$P_1$ or $P_2$. The points in the interval $[x_1, x_2]$ are a subset of the
points stored along $P_1$ and $P_2$ and the set of all points in $T_1, \ldots,
T_h$. Each tree $T_i$ is a heap on the points it stores. In particular, the
point with maximum $y$-coordinate in the range $[x_1, x_2]$ is stored along
$P_1$, along $P_2$ or at the root of some subtree~$T_i$.

To answer a range top-$k$ query with interval $[x_1, x_2]$, we insert all points
stored along $P_1$ or $P_2$ with $x$-coordinates in the interval $[x_1, x_2]$,
as well as the point stored at the root of each subtree $T_i$ into a binary heap
$Q$. In total, we insert $O(\lg n)$ points into $Q$. We now repeat the following
operation $k$ times. We retrieve the point $p$ with maximum $y$-coordinate from
$Q$ using a \textsc{DeleteMax} operation. Let $p'$ be the point in $[x_1, x_2]$
with the next lower $y$-coordinate. If $p$ is stored on $P_1$ or $P_2$, then
$p'$ is one of the remaining points stored along $P_1$, stored along $P_2$, or
stored at the roots of the subtrees $T_1, \ldots, T_h$.  If $p$ was stored in
some subtree $T_i$, then $p'$ is one of the remaining points in $Q$ or it is one
of the two points stored at the children of the node of $T_i$ where $p$ was
stored. We insert these two points into $Q$ to ensure that the next
\textsc{DeleteMax} operation retrieves~$p'$. This process continues until we
have retrieved the $k$ points with maximum $y$-coordinates among all points in
the $x$-range $[x_1, x_2]$.

Traversing $P_1$ and $P_2$ and constructing $Q$ takes $O(\lg n)$ time because
these two paths have length $O(\lg n)$ and a binary heap can be built in linear
time in the number of elements it stores initially. A \textsc{DeleteMax}
operation takes $O(\lg m)$ time, where $m$ is the number of elements in the
heap. Initially, $m = O(\lg n)$. For every point retrieved from $Q$, we insert
at most two points into $Q$. Since we retrieve $k$ points from $Q$, $m$ never
exceeds $k + O(\lg n)$, so the cost of each \textsc{DeleteMax} operation is
$O(\lg(k + \lg n)) = O(\lg\lg n + \lg k)$. The cost of the entire range
top-$k$ query is thus $O(\lg n + k(\lg \lg n + \lg k)) = O(k \lg n)$.

\subsection{Range Sum Data Structure}

\label{sec:range-sum-ds}

A range sum data structure maintains a set of pairs $(x, w)$ sorted by their
$x$-coordinates. The query we are interested in is the following: Given an
interval $[x_1, x_2]$, report the sum of the $w$-coordinates of all pairs $(x,
w)$ with $x_1 \le x \le x_2$.

To support this operation, a range sum data structure stores the pairs in a
binary search tree $T$ on their $x$-coordinates. Each node of $T$ stores the sum
of the $w$-coordinates of all pairs associated with its descendant leaves. To
answer a range sum query with interval $[x_1, x_2]$, we traverse the two paths
$P_1$ and $P_2$ corresponding to $x_1$ and $x_2$ and sum the totals stored at
the roots of the subtrees between $P_1$ and $P_2$. We also add the
$w$-coordinates of the pairs associated with the leaves at the bottom of $P_1$
and $P_2$ if these pairs have $x$-coordinates in the query range.  Since $P_1$
and $P_2$ have length $O(\lg n)$, there are $O(\lg n)$ values to be summed,
which takes $O(\lg n)$ time.

A range sum data structure also supports insertions and deletions in $O(\lg n)$
time. The details are straightforward.

Note that the maintenance of a range sum data structure, and the answering of
range sum queries only requires the $w$-coordinates of the pairs stored in the
data structure to be drawn from a semigroup (see \cref{sec:semigroup}).

\subsection{Range Tree}

\label{sec:range-tree}

A $d$-dimensional range tree \cite{bentleyDecomposableSearchingProblems1979} is
a tree $T$ to store a set of points $S \subset \RR^d$ and is defined as follows:
If $d = 1$, then $T$ is a binary search tree on $P$.  If $d > 1$, then refer to
the coordinates of the points to be stored in $T$ as their $x_i$-coordinates,
for $1 \le i \le d$. $T$ is a binary search tree on the $x_1$-coordinates of the
points in $S$. For each internal node $v$ of $T$, let $S_v$ be the subset of
points stored at descendant leaves of $v$. Then we associate with each node $v$
a $(d-1)$-dimensional range tree on the $x_2$ through $x_d$-coordinates of the
points in $S_v$.

In our discussion of range trees, we refer to the tree on the $x_1$-coordinates
as the \emph{level-$1$ tree}, to the trees associated with the nodes of the
level-$1$ tree as \emph{level-$2$ trees}, to the trees associated with their
nodes as \emph{level-$3$ trees}, and so on. In general, the level-$i$ tree is a
binary search tree on the $x_i$-coordinates of the points it stores.

Bentley \cite{bentleyDecomposableSearchingProblems1979} showed that a
$d$-dimensional range tree uses $O\bigl(n \lg^{d-1} n\bigr)$ space and can be
constructed in $O\bigl(n \lg^{d-1} n\bigr)$ time.  Given a $d$-dimensional query
range $[l_1, r_1] \times [l_2, r_2] \times \cdots \times [l_d, r_d]$, all points
in $S$ that belong to this range can be reported in $O\bigl(\lg^d n + k\bigr)$
time, where $k$ is the number of reported points. Willard and Lueker
\cite{willardAddingRangeRestriction1985} showed that a $d$-dimensionsal range
tree supports insertions and deletions in $O\bigl(\lg^d n\bigr)$ amortized time.
Their proof uses only that binary search trees support insertions and deletions
in $O(\lg n)$ time and can be built in linear time, provided that the points to
be stored in the tree are already sorted.

\subsection{Combinations}

\label{sec:combinations}

As building blocks for our SOLE data structures, we need two straightforward
combinations of priority search trees, range sum data structures, and range
trees.

The first is a ``range sum priority search tree.'' It is a priority search tree
$T$ on a set of pairs $(x,w) \in \RR \times S$, where $S$ is any totally ordered
semigroup. The $w$-components of these pairs, called \emph{weights}, are treated
as the $y$-coordinates.  Each node $v$ of $T$ also stores the total weight of
all the pairs stored at $v$ and its descendants in $T$. This data structure can
clearly be maintained in $O(\lg n)$ time per insertion and deletion and supports
both range sum queries and reporting the $k$ pairs with maximum weight among
those whose $x$-coordinates are in some query range $[x_1, x_2]$, in $O(\lg n)$
time and $O(k \lg n)$ time, respectively.

The second combination we need is a ``$d$-dimensional range sum priority search
tree.''  It stores a set of $(d+1)$-tuples $(x_1, \ldots, x_d, w) \in \RR^d
\times S$, where $S$ is once again any totally ordered semigroup.  The
underlying structure is a $d$-dimensional range tree on the first $d$
coordinates of these tuples. Every level-$i$ tree, at any level $1 \le i \le d$,
is organized as a ``range sum priority search tree'', using the $w$-coordinates
of all tuples as the priorities.  In other words, the points associated with the
leaves of each such tree are propagated upwards as in a priority search tree, to
turn the tree into a heap on the $w$-coordinates of these points. Each node of
such a tree also stores the total weight of all the tuples associated with its
descendant leaves. Since a range sum priority search tree supports updates in
$O(\lg n)$ time and can be built in linear time, given the points to be stored
in it in sorted order, this ``$d$-dimensional range sum priority search tree''
supports insertions and deletions in amortized $O\bigl(\lg^d n\bigr)$ time,
using the same approach described by Willard and Lueker
\cite{willardAddingRangeRestriction1985}.

Given such a tree, we can now support $d$-dimensional range sum and
$d$-dimen\-sional range top-$k$ queries. A $d$-dimensional range sum query with
query range $[l_1, r_1] \times [l_2, r_2] \times \cdots \times [l_d, r_d]$
traverses the level-1 tree to find all subtrees between the two paths $P_1$ and
$P_2$ corresponding to $l_1$ and $r_1$. It recursively answers a
$(d-1)$-dimensional range sum query on the $(d-1)$-dimensional range trees
associated with the roots of these subtrees. The result is the sum of the totals
returned by these queries plus the weight of those tuples stored along $P_1$ and
$P_2$ that fall within the given query range. A simple inductive argument shows
that the cost of such a query is $O\bigl(\lg^d n\bigr)$.

A $d$-dimensional range top-$k$ with query range $q = [l_1, r_1] \times [l_2,
r_2] \times \cdots \times [l_d, r_d]$ traverses the paths $P_1$ and $P_2$
corresponding to $l_1$ and $r_1$ in the level-1 to identify the subtrees between
these two paths. It recursively answers $(d-1)$-dimensional range top-$k$
queries on the $(d-1)$-dimensional range trees associated with the roots of
these subtrees. The top $k$ tuples in $q$ are among the $O(k \log n)$ elements
returned by these recursive queries and those tuples stored along $P_1$ and
$P_2$ that fall within the query range $q$. These top $k$ tuples can now be
found in $O(k \lg n)$ time using linear selection
\cite{blumTimeBoundsSelection1973}. A simple inductive argument shows that the
cost of a $d$-dimensional range top-$k$ query is thus $O\bigl(k \lg^d n\bigr)$.

\end{document}